\documentclass[10pt]{article}
\usepackage{citesort,multicol,epsfig}
\usepackage{overcite}
\begin{document}

\title{Nucleation of the crystalline phase of proteins in the presence
of semidilute non-adsorbing polymer}

\author{{\bf Richard P. Sear}\\
~\\
Department of Physics, University of Surrey\\
Guildford, Surrey GU2 7XH, United Kingdom\\
email: r.sear@surrey.ac.uk}

\maketitle

\begin{abstract}
Starting from a protein solution which is metastable with respect
to the crystalline phase, the effect of adding semidilute non-adsorbing
polymer is considered. It is found to increase the chemical potential of the
protein by a few tenths of kT, which may be enough to lower the
barrier to nucleation of the crystalline phase by enough to allow
crystallisation. It is also shown that assuming that the polymer induces a
pairwise additive attraction leads to qualitatively incorrect results.
\end{abstract}

\begin{multicols}{2}
\section{Introduction}

Crystallising proteins from dilute solution is in general rather difficult,
yet the crystalline form is required in order to perform
X-ray crystallography \cite{chayen98,durbin96}. X-ray crystallography is
currently the best method of determining the all-important structure
of a protein in its native state. This has lead to considerable
interest in how proteins crystallise; see Ref. \citen{piazza00}
for a recent review from a colloidal physics perspective
and Refs. \citen{talanquer98,haas00,dixit00,pini00,search}
for recent work.
In a dilute solution there may be a large free energy barrier
to nucleation of the crystalline phase \cite{chayen98,piazza00,searxxx},
which prevents nucleation on accessible time scales.
Here, we consider a generic model protein which in the {\em absence} of
polymer already has coexistence between a dilute solution and a
crystalline phase. We determine
how the barrier to homogeneous nucleation of the
crystalline phase changes when non-adsorbing
polymer coils which are larger than the protein
molecules are added to the protein solution. 
We find that the polymer lowers the nucleation barrier. This is to be
expected as it is well-known that non-adsorbing polymer induces an effective
attraction between the protein molecules, and an attraction will
favour the dense crystalline phase. However, what is much less
common, is that for
large polymer coils it is essential to account for the fact that the
polymer-induced attraction is strongly non-pairwise
additive. A naive calculation using in the crystalline phase,
the effective pair potential in
dilute phase of protein molecules, greatly
overestimates the effect of the polymer.

Protein molecules are typically a few nms across and the root-mean-square
end-to-end distance of a flexible polymer can easily be several times this.
The protein--protein interaction
consists of a repulsive core of diameter $D$ and an attraction
which is sufficiently strong to cause the
equilibrium behaviour to be coexistence between a dense crystal and a dilute
fluid phase. We do not need to specify the potential explicitly as we will
require only the fluid-crystal interfacial tension $\gamma_0$ and the
difference in chemical potential between that of the crystal and that
of the fluid, $\Delta\mu_0$. We use the subscript 0 to indicate
quantities before polymer is added. Our results should therefore apply
to a very wide range of proteins despite potentially very large
variations in the structure and interactions
of different proteins \cite{durbin96}.
The polymer is non-adsorbing:
the interaction between the protein molecule and a monomer is
purely repulsive.

There has been extensive work done on systems of small
colloidal particles immersed in polymer solutions of larger
polymer coils which do not adsorb on the particles
\cite{smallnote}. Scaling theory for these systems was
pioneered by de Gennes \cite{degennes79}, see also Refs.
\citen{odijk96,odijk00,sear97,sear98}.
Field-theoretic methods have yielded results which
are exact for small hard particles and ideal polymers \cite{eisenriegler96},
as well as approximate results for polymers in good solvents
\cite{hanke99,eisenriegler00}.
Work has also been done using integral equations, see
Ref. \citen{fuchs00} and
references therein.
It is well understood
that non-adsorbing polymer increases the chemical potential of particles and
induces an attraction between these particles which is not pairwise
additive: the interaction of
three close-by protein molecules is not the sum of three
pair potentials \cite{meijer94,odijk97,searp01,hanke99}.
However, most of this work (exceptions are Refs. \citen{odijk97,searp01})
only considered a few particles, two or three,
whereas a nucleus of a crystalline phase contains 10s of tightly-packed
particles. Because of this, the deviations from pairwise additivity observed 
here are much stronger than found in previous work.

The next section briefly outlines classical nucleation theory for
the solution before polymer is added. Section 3 then describes the
effect of polymer and section 4 is a conclusion.

\section{Classical nucleation theory}

The interaction between protein molecules
is taken to include a steeply repulsive core and an
attraction. This attraction may be isotropic or
highly anisotropic or a mixture of the two, if isotropic it may
or may not be short-ranged; all we require is that
it render a dilute solution of the protein metastable with respect
to a crystalline phase of the protein. By metastable we mean that the
fluid phase is not the equilibrium phase, it does not correspond
to the absolute minimum in the appropriate free energy, but the fluid
phase is dynamically stable with respect to crystallisation for times
much larger than the characteristic time of the solution.
See Ref. \citen{debenedetti} for a general introduction to the properties
and behaviour of metastable fluids.

Classical nucleation theory \cite{debenedetti,gunton,chaikin}
assumes that the nucleus of the
crystalline phase has a free energy $\Delta F$ which is the sum
of a bulk term and surface term.
See the book of Debenedetti \cite{debenedetti} for an excellent
introduction to classical nucleation theory.
The protein molecules are modeled
by spherical particles of diameter $D$.
The bulk term is equal to the number of molecules in the nucleus, $n$,
times the chemical potential difference $\Delta\mu=\mu_x-\mu$,
where $\mu_x$ is the chemical potential of the crystalline phase
and $\mu$ is the chemical potential
of the fluid phase which contains the nucleus. The fluid phase
we are considering is not the true equilibrium phase, nucleation occurs
from a fluid which is at a higher chemical potential than the crystal. Thus
$\Delta\mu$ is negative.
The surface term is the surface area of the nucleus
times the surface tension $\gamma$ of the bulk
interface between the coexisting crystalline and fluid phases.
The surface area of the nucleus is obtained by assuming that the
nucleus is a sphere of radius $R$ which is related to $n$
by assuming that the density of spheres within the nucleus
is equal to the bulk density of the crystalline phase, taken
to be $D^{-3}$: the density of a close-packed simple cubic lattice.
Then $n=(4/3)\pi R^3D^{-3}$. Putting this all together,
\begin{equation}
\Delta F = \frac{4}{3}\pi R^3D^{-3}\Delta\mu + 4\pi R^2\gamma
\label{df}
\end{equation}
The first term in $\Delta F$ is the bulk term,
which is negative and decreases linearly with $n$, and the
second term is the surface term
which is positive and increases as $n^{2/3}$. Thus
$\Delta F$ passes through
a maximum, denoted by $\Delta F^*$, at some value of $R$,
denoted by $R^*$. This maximum is the top of the free energy
barrier to nucleation.
The nucleus with radius $R^*$ and $n^*$ spheres is called
the critical nucleus.
Taking the derivative of Eq. (\ref{df}) with respect to $R$ and
equating to zero,
\begin{equation}
R^* = \frac{2\gamma D^3}{|\Delta\mu|},
\label{ncnt}
\end{equation}
and inserting this value of $n$ in Eq. (\ref{df})
\begin{equation}
\Delta F^* = \frac{16\pi\gamma^3D^6}
{3\Delta\mu^2}.
\label{cnt}
\end{equation}
Nucleation is an activated process with $\Delta F^*$ as the barrier
height and so the nucleation rate varies as 
$\exp(-\Delta F^*/kT)$, where $k$ and $T$ are Boltzmann's constant and
the temperature $T$, respectively.
To increase the rate of nucleation of the
crystalline phase of the protein we need to reduce the barrier
$\Delta F^*$. In the absence of polymer $\Delta F^*$ and $R^*$
are obtained from Eqs. (\ref{ncnt}) and (\ref{cnt}) by setting
$\Delta\mu=\Delta\mu_0$ and $\gamma=\gamma_0$.

\section{Nucleation in the presence of semidilute polymer}

\begin{figure}
\begin{center}
\caption{
\lineskip 2pt
\lineskiplimit 2pt
A schematic of a crystalline nucleus in a polymer solution.
The black discs represent the protein molecules of the nucleus,
and the curves are polymer molecules.
\label{figschem}
}
\vspace*{0.2in}
\epsfig{file=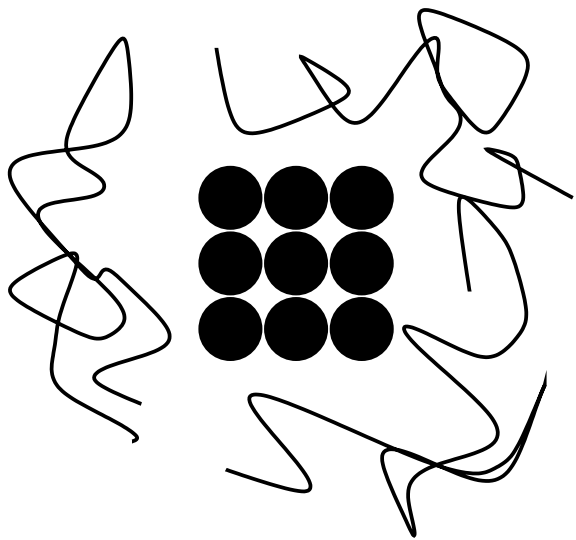,width=3.0in}
\end{center}
\end{figure}

Now we consider
a solution of both protein molecules and flexible polymer molecules.
The
interaction between a protein molecule and a monomer is taken
to be strictly repulsive. Figure \ref{figschem} is a schematic
of the nucleus in a solution of polymer.
We remark that this simple model is
not adequate for all polymers, for example it is not
adequate for PEG (poly(ethylene glycol)).
We consider the case where the polymer molecules are larger
than the protein molecules and the polymer solution is semidilute.
As the protein molecules are
only a few nms in diameter this regime is easily accessible in experiment.
In the other limit, where the polymer's radius is smaller than that of the
colloidal particles, the polymer induces a short-ranged
depletion attraction between the particles. Such attractions
are to a good approximation pairwise additive \cite{smallnote}
and therefore straightforward
to treat. See
Refs. \citen{meijer94,dijkstra99,dijkstra00,louis00,louis01}
for work where the polymer is at most as large as the colloid.

See the book of de Gennes \cite{degennes} for an introduction to semidilute
polymer solutions. They are characterised by a correlation length
$\xi$ which scales as $c_M^{-3/4}$, where $c_M$ is the monomer
concentration. The osmotic pressure exerted by the polymer
$\Pi\sim kT\xi^{-3}$.

The crystalline phase of the protein is dense, the gaps between protein
molecules available to polymer are around $0.4D$ or less across.
Forcing a polymer
molecule into a tube of diameter $0.4D$ costs roughly $kT$ in free energy
per $0.4D$ of the tube's length as the polymer collides with the
walls of the tube at intervals of roughly its diameter. Thus, the
free energy density inside such
a crystal, either a bulk crystal or a crystalline
nucleus, will be
of order $kT$ per $0.06D^3$ volume whereas in solution it is $kT$
per $\xi^3$ volume. For $\xi$ larger than $D$, which we assume throughout,
the free energy density inside a crystal is so much higher than outside
that to a good approximation there is no polymer inside the crystal.
This agrees with calculations using the theory of Ref. \citen{searp01}.
So, as the crystalline nucleus is impermeable to polymer it interacts
with the polymer, as a hard particle of radius $R$.

The interaction $w$ of a hard spherical particle of
diameter $D$ with a semidilute solution of polymer
in a good solvent with a correlation length $\xi$ is
\cite{degennes79,odijk96,odijk00,sear97}
\begin{equation}
w\sim kT \left(\frac{D}{\xi}\right)^{4/3}~~~~~~D\ll\xi ,
\label{wsmall}
\end{equation}
where we took the exponent $\nu$ to have its Flory value of $3/5$
\cite{degennes}.
$w$ is the work done (= the difference in excess chemical potential)
in taking a particle
from a pure solvent and inserting it into the polymer solution.
This assumes that the
interaction between a protein and the monomers of a polymer is purely
repulsive and that there is not much more than one protein molecule
per $\xi^3$ volume.
The notation $\sim$ indicates that we have neglected a coefficient
of order unity; here we derive only the scaling behaviour of the
interactions with respect to the relevant length scales.
Equation (\ref{wsmall}) can be derived in a couple
of ways, see Refs. \citen{degennes79,odijk96,odijk00,sear97}.
One way is to note that
when $D\ll\xi$ the interaction between a particle and polymer must
scale as the density of monomers, which scales as $\xi^{-4/3}$
when $\nu=3/5$. As $w/kT$ is dimensionless, we require that $\xi^{-4/3}$
appears as the ratio $(D/\xi)^{4/3}$ which gives Eq. (\ref{wsmall}).


Straightaway we can derive an approximation for the
contribution of polymer to the difference in excess chemical
potential $\Delta\mu$ between the dense crystalline
phase of the protein, and the semidilute polymer solution.
It is
\begin{equation}
\Delta\mu\sim\Delta\mu_0-kT\left(\frac{D}{\xi}\right)^{4/3}+\Pi D^3 ,
\label{d1}
\end{equation}
where the second term on the right hand side
is the change in free energy of the polymer
when a single protein molecule is removed from the semidilute
polymer solution and the last term is the change in free energy
of this solution when it is compressed by the crystalline phase of the
protein expanding by the volume of one protein.
The osmotic pressure of a polymer solution $\Pi$ is related to $\xi$ by
$\Pi\sim kT\xi^{-3}$. Thus, the
last term is equal to $kT(D/\xi)^3$ and so is
smaller than the second term, indeed it is of order terms we have
dropped and so we drop it and obtain our final expression for
$\Delta \mu$ in the presence of polymer,
\begin{equation}
\Delta\mu\sim\Delta\mu_0-kT\left(\frac{D}{\xi}\right)^{4/3}.
\label{dmup}
\end{equation}

For a large nucleus, radius $R\gg\xi$, the width of the
nucleus--polymer-solution interface, $\xi$,
is small and so the capillarity approximation
which underlies classical nucleation theory still holds: $\Delta F$
is still a sum of bulk and interfacial terms. The bulk term is given
by the $\Delta\mu$ of Eq. (\ref{dmup}) times the number of protein
molecules in the nucleus.
In the presence of semidilute polymer the
the interfacial tension of a flat interface, $\gamma$ is \cite{joanny79},
\begin{equation}
\gamma \sim \gamma_0 + \frac{kT}{\xi^2},
\label{gamp}
\end{equation}
where $\gamma_0$ is the interfacial tension in the absence of polymer.
Bearing in mind that $\gamma_0$ will be at least of
order $kT/D^2$ we see that the fractional modification
to $\gamma_0$ due to polymer is of order $(D/\xi)^2$.
For a $\Delta\mu_0$ of order $kT$, the fractional change
to $\Delta\mu$ due to polymer is, see Eq. (\ref{dmup}), of order
$(D/\xi)^{4/3}$. Within our crude scaling theory we only consider
leading order contributions and so neglect the contribution of the
polymer to $\gamma$, leaving the only contribution of polymer as the
change in $\Delta\mu$ given by Eq. (\ref{dmup}).

In the opposite limit for $R$, $R\ll \xi$, then the
contribution of polymer to the free energy of the nucleus is no longer
the sum of bulk and interfacial terms,
i.e., it is not the sum of terms scaling as $R^3$ and as
$R^2$.
For $R\ll\xi$ the contribution of the polymer to
the free energy of formation of the nucleus $\Delta F^*$ is the
sum of two terms: the first is the
number of protein molecules in the nucleus times minus the increase in
chemical potential in solution due to polymer,
Eq. (\ref{wsmall}), and the second is
the free energy cost of inserting the nucleus into
the polymer solution. This second term is the
free energy cost due to the nucleus excluding polymer from a sphere of
radius $R$.
Because the polymer cannot penetrate
into the nucleus it interacts with the polymer as a single particle, and so
this free energy cost is just Eq. (\ref{wsmall}) with
$D$ replaced by $2R$. Adding the two terms together we have the
contribution of polymer to $\Delta F$,
\begin{equation}
\Delta F=\Delta F_0-nkT\left(\frac{D}{\xi}\right)^{4/3}+
kT\left(\frac{2R}{\xi}\right)^{4/3}~~~~~R<\xi ,
\label{smallr}
\end{equation}
where $\Delta F_0$ is the free energy of formation of the nucleus before
polymer is added.
For $R$ a few times $D$
and $R,D\ll\xi$ the second term in Eq.
(\ref{smallr}) dominates the last term. So, we neglect the last term
in Eq. (\ref{smallr}).
Then as the remaining term is linear in $n$ it is in effect
a shift in $\Delta\mu$ and indeed is the same shift in $\Delta\mu$
we found in the opposite limit, $R\gg\xi$. Thus
we conclude that as for $R\gg\xi$ the leading
order contribution of semidilute polymer to the nucleation barrier
is simply to shift $\Delta\mu$ by the amount given in Eq. (\ref{dmup}).

So, we have shown that in both limits, $R\gg\xi$ and $R\ll\xi$,
the effect of adding polymer in the semidilute regime is to shift the chemical
potential difference $\Delta\mu$ by an amount of order $-(D/\xi)^{4/3}kT$,
which will be a few tenths of $kT$. Having shown that it holds in both limits
we assume that it also holds for $R/\xi=O(1)$ as well.
The shift varies as $\xi^{-4/3}$ and so increases linearly with
the density of polymer. The size of the critical nucleus also
decreases, Eq. (\ref{ncnt}).
Thus, if decreasing
$\Delta\mu$ by a few tenths of $kT$ is enough to reduce the
nucleation barrier sufficiently to make the nucleation
rate significant then adding semidilute polymer is a viable way
of inducing nucleation of the crystalline phase of small colloidal
particles such as proteins and micelles.

\subsection{Comparison with assumption of pairwise additivity}

A pair of particles in a semidilute polymer solution
separated by a distance $x$ of order $\xi$ or
less feel a polymer-induced attraction towards each other,
which we denote by $w_2$ \cite{sear97,hanke99,eisenriegler00}.
For small particles, $D\ll \xi$, and for separations $r$
much smaller than the correlation length
\cite{sear98,hanke99}
\begin{equation}
w_2\sim - kT
\left(\frac{D}{\xi}\right)^{4/3}\left(\frac{D}{r}\right)^{4/3}
~~~~~~~ r \ll \xi .
\label{hanke}
\end{equation}
Naively it
might be thought that the contribution of polymer to the free energy
change $\Delta \mu$ could be estimated using this pair attraction
plus a mean-field approximation, which would correspond to the
approximation
\begin{equation}
\Delta \mu\sim\Delta \mu_0 + D^{-3}\int_{r>D}{\rm d}{\bf r} w_2(r)
~~~~~~\mbox{wrong},
\label{mf1}
\end{equation}
which is just the integral over the number density, $D^{-3}$ in the
crystalline phase, times the potential.
As usual with a mean-field approximation, correlations in the positions
of particles are neglected and one protein is considered to interact
with surrounding particles which are at the mean density.
The above integral is $D^{-3}$ times the contribution of polymer
to the second virial coefficient. See Eisenriegler \cite{eisenriegler00}
for a more accurate treatment of the contribution of polymer to
the second virial coefficient.
So, inserting Eq. (\ref{hanke}), into Eq. (\ref{mf1}), we obtain
\begin{equation}
\Delta \mu - \Delta\mu_0\sim - kT
\left(\frac{\xi}{D}\right)^{1/3}
~~~~~~\mbox{wrong}.
\label{mf}
\end{equation}
If we compare Eqs. (\ref{dmup}) and (\ref{mf}) we see that
the assumption of a pairwise additive potential plus the mean-field
approximation predicts the wrong scaling with $(D/\xi)$ of the polymer
contribution to $\Delta \mu$.
It greatly overestimates,
by a factor of order $(\xi/D)^{5/3}$,
the effect of adding polymer whose
correlation length is large with respect to the protein diameter.
Naively using the pair potential between an isolated pair of particles,
(a pair of particles with no others within a distance $\xi$ of them),
as the pair interaction within a dense, number density $\gg \xi^{-3}$
(here the crystal),
is qualitatively wrong.

\section{Conclusion}

We have estimated the change in the barrier to nucleation
when polymer is added to
a metastable dilute protein solution. The polymer
is semidilute, and has a correlation length $\xi$
larger than the diameter of the
protein $D$; the polymer does not adsorb onto the surface of the protein.
When polymer is added, the barrier $\Delta F^*$ is reduced
due to the increase in the chemical potential of the protein in solution,
possibly by enough to allow nucleation.
However, the effect is not large, $\Delta\mu$ decreases by
only a few tenths of $kT$. Also, a naive use of the polymer induced
attraction between a pair of proteins yields results which are qualitatively
wrong: they grossly overestimate the effect of adding polymer.
This does not mean that an effective potential approach to the effect of
polymer cannot reliably estimate the effect of large polymer molecules, just
that this effective potential is many-body in nature and cannot
simply be approximated by just a pair potential. This conclusion
also applies to the phase behaviour of a system of hard-sphere-like colloidal
particles and larger polymer molecules \cite{searp01}.

Finally, we note that the second virial coefficient $B_2$ is often used to
measure the strength of the attractions in a protein solution.
$B_2$ is given by
\begin{eqnarray}
B_2 &=& B_{2,0}+\frac{1}{2}\int_{r>D}{\rm d}{\bf r} \frac{w_2(r)}{kT}
~~~~~~~~w_2\ll kT
\label{b2}\\
&\sim &B_{2,0} - D^{8/3}\xi^{1/3},
\end{eqnarray}
where $B_{2,0}$ is the second virial coefficient in the absence
of polymer and we have assumed not only that $w_2/kT$ is small but that
$B_{2,0}$ is dominated by interactions with a range $\ll\xi$: then
the second virial coefficient is the sum of the two terms in Eq. (\ref{b2}).
So as the polymer density {\em increases},
$\xi$ decreases and the contribution of polymer to $B_2$
{\em  decreases}. But
from Eq. (\ref{dmup}) we see that as the polymer density increases it has
a larger and larger effect on the nucleation barrier. So, the
effect of polymer on $B_2$ is {\em anti}-correlated with its effect on
$\Delta F^*$, i.e., when one goes up the other goes down,
not correlated as we might naively have expected.


\end{multicols}


\begin{thebibliography}{99}

\bibitem{chayen98} N. Chayen and J. Helliwell,
Physics World {\bf 11}, 43 (May) (1998).

\bibitem{durbin96} S. D. Durbin and G. Feher,
Ann. Rev. Phys. Chem. {\bf 47}, 171 (1996).

\bibitem{piazza00} R. Piazza,
Curr. Opinion Coll. Int. Sci. {\bf 5}, 38 (2000).

\bibitem{talanquer98} V. Talanquer and D. W. Oxtoby,
J. Chem. Phys. {\bf 109}, 223 (1998).

\bibitem{haas00} C. Haas and J. Drenth,
J. Phys. Chem. B {\bf 104}, 368 (2000).

\bibitem{dixit00} N. M. Dixit and C. F. Zukoski,
J. Coll. Int. Sci. {\bf 228}, 359 (2000).

\bibitem{pini00} D. Pini, G. Jialin, A. Parola and L. Reatto,
Chem. Phys. Lett. {\bf 327}, 209 (2000).

\bibitem{search} R. P. Sear, J. Chem. Phys, {\bf 114}, 3170 (2001).

\bibitem{searxxx} R. P. Sear,
cond-mat/9912199.

\bibitem{smallnote}
As emphasised by Evans and coworkers, see for
example Refs. \citen{dijkstra99,dijkstra00}, the interaction is only strictly
pairwise additive if the polymer is both ideal and does not interact
with a particle more than $\simeq 0.15$ times its radius
away. However, if the polymer
is larger than this but still small with respect to the protein molecule,
the potential will be close to pairwise additive. Then the
simultaneous interaction of a polymer molecule with more than two
protein molecules (which is the origin of deviations from pairwise
additivity) will be weak. In the other limit, which we consider here, where
the polymer is much larger than the protein, the polymer molecule
can potentially interact with {\em many} protein molecules, inducing
an effective interaction which is very far from
being pairwise additive.

\bibitem{degennes79} P. G. de Gennes, C. R. Acad. Sci. Paris B {\bf 288},
359 (1979).

\bibitem{odijk96} T. Odijk, Macromolecules {\bf 29}, 1842 (1996).

\bibitem{odijk00} T. Odijk, Physica A {\bf 278}, 347 (2000).

\bibitem{sear97} R. P. Sear, Phys. Rev. E {\bf 56}, 4463 (1997).

\bibitem{sear98} R. P. Sear, Eur. Phys. J. B {\bf 1}, 313 (1998).

\bibitem{eisenriegler96} E. Eisenriegler, A. Hanke and S. Dietrich,
Phys. Rev. E {\bf 54}, 1134 (1996).

\bibitem{hanke99} A. Hanke, E. Eisenriegler and S. Dietrich,
Phys. Rev. E {\bf 59}, 6853 (1999).

\bibitem{eisenriegler00} E. Eisenriegler,
J. Chem. Phys. {\bf 113}, 5091 (2000).

\bibitem{fuchs00} M. Fuchs and K. S. Schweizer,
Europhys. Lett. {\bf 51}, 621 (2000).

\bibitem{searp01} R. P. Sear,
cond-mat/0012362.

\bibitem{odijk97} T. Odijk, J. Chem. Phys. {\bf 106}, 3402 (1997).

\bibitem{meijer94} E. J. Meijer and D. Frenkel,
J. Chem. Phys. {\bf 100}, 6873 (1994).

\bibitem{debenedetti} P. G. Debenedetti,
{\it Metastable Liquids}
(Princeton University Press, Princeton, 1996).

\bibitem{gunton} J. D. Gunton, M. San Miguel and P. S. Sahni,
in {\it Phase Transitions} volume 8, edited by C. Domb and J. L. Lebowitz
(Academic Press, London, 1983).

\bibitem{chaikin} P. M. Chaikin and T. C. Lubensky,
{\it Principles of Condensed Matter Physics}
(Cambridge University Press, Cambridge, 1995).

\bibitem{dijkstra99} M. Dijkstra, J. M. Brader and R. Evans,
J. Phys. Cond. Mat. {\bf 11}, 10079 (1999).

\bibitem{dijkstra00} M. Dijkstra, R. van Roij and R. Evans,
J. Chem. Phys. {\bf 113}, 4799 (2000).

\bibitem{louis00} A. A. Louis, P. G. Bolhuis, J. P. Hansen, E. J. Meijer,
Phys. Rev. Lett. {\bf 85}, 2522 (2000).

\bibitem{louis01} A. A. Louis,
cond-mat/0102220.

\bibitem{degennes} P. G. de Gennes,
{\it Scaling Concepts in Polymer Physics}
(Cornell University Press, Ithaca, 1979).

\bibitem{joanny79} J. F. Joanny, L. Leibler and P.-G. de Gennes,
J. Polymer Sci. Polymer Phys. Ed. {\bf 17}, 1073 (1979).




\end{thebibliography}
\end{document}